\documentclass[aps,prc,reprint,superscriptaddress,showpacs,nofootinbib]{revtex4-1}
\usepackage{graphicx,psfrag}% Include figure files
\usepackage{dcolumn}% Align table columns on decimal point
\usepackage{bm}% bold math
\usepackage{subfigure}
\usepackage{amsmath}
\usepackage{amssymb}
\usepackage{color}
\usepackage{framed}
\usepackage[normalem]{ulem}

\begin{document}

\title{Formation of $\eta'(958)$-mesic nuclei by ($p,d$) reaction}

\author{Hideko Nagahiro}
   \affiliation{Department of Physics, Nara Women's University, Nara,
   630-8506, Japan}

\author{Daisuke Jido}
   \affiliation{Yukawa Institute for Theoretical Physics, 
Kyoto University, Kyoto 606-8502, Japan}
   \affiliation{J-PARC Branch, KEK Theory Center, Institute of Particle
   and Nuclear Studies, High Energy Accelerator Research Organization
   (KEK), Ibaraki, 319-1106, Japan}

\author{Hiroyuki Fujioka}
   \affiliation{Division of Physics and Astronomy, Kyoto University,
   Kyoto, 606-8502, Japan} 

\author{Kenta Itahashi}
   \affiliation{Nishina Center for Accelerator-Based Science, RIKEN, 
2-1 Hirosawa, Wako, 351-0198 Saitama, Japan}

\author{Satoru Hirenzaki}
   \affiliation{Department of Physics, Nara Women's University, Nara, 630-8506, Japan}
\begin{abstract}
{We {calculate} theoretically the formation spectra of
 $\eta'(958)$-nucleus systems in the $(p,d)$ reaction {for the
 investigation of}
 the {in-medium modification of the $\eta'$ mass}.  We show
 the comprehensive numerical calculations based on a simple form of the
 {$\eta'$} optical potential {in nuclei} with various potential
 depths.  We conclude that {one finds}
 an evidence of possible attractive interaction between $\eta'$
 and nucleus {as} peak structure appearing around the $\eta'$
 threshold in light nuclei such as $^{11}$C when the attractive
 potential is stronger than 100 MeV and the absorption width is of order
 of 40 MeV or less.  
{Spectroscopy} of the $(p,d)$ reaction
 is expected to be performed {experimentally} at {existing
 facilities, such as} GSI.  We also estimate the contributions 
 from the $\omega$ and $\phi$ mesons, which have {masses close} to the
 $\eta'$ meson, {concluding} that the observation of
 the peak structure of the $\eta'$-mesic nuclei {is not disturbed}
 although their contributions may not be small.}
\end{abstract}
%\pacs{21.85.+d, 21.65.Jk, 12.39.Fe, 14.20.Gk, 14.40.Aq, 25.80.Hp}
\maketitle

\section{Introduction}
\label{intro}
The $\eta'(958)$ meson is an interesting and important particle because
of its exceptionally large{r} mass and connection to the $U_A(1)$
problem~\cite{Weinberg:1975ui}. 
According to the symmetry pattern of the quark sector in QCD, the
$\eta'$ meson would be one of the Nambu-Goldstone bosons associated with
the spontaneous breakdown of the $U(3)_L\times U(3)_R$ chiral symmetry
to the $U_V(3)$ flavor symmetry.  In the real world, however, gluon
dynamics plays an important role, and the $\eta'$ meson acquires its
peculiarly larger mass than those of the other pseudoscalar mesons,
$\pi$, $K$, and $\eta$ through the quantum anomaly effect of
non-perturbative gluon dynamics~\cite{Witten:1979vv,Veneziano:1979ec}
which induces the non-trivial vacuum structure of
QCD~\cite{'tHooft:1976up}.  The mass generation of the $\eta'$ meson is
considered to be 
a result of the interplay of quark symmetry and gluon dynamics.  
{
The $\eta'$ meson at finite density has been discussed for long
time~\cite{Pisarski:1983ms,Kapusta:1995ww,Bass:2005hn,Bernard:1987sg,Tsushima:1998qp} 
and the possibility of the $\eta'$ mesic nuclei formation has been first
investigated in Ref.~\cite{Nagahiro:2004qz}.}

Recently, there have been two important developments in
theoretical~\cite{Jido:2011pq} and experimental~\cite{Itahashi:2012ab,*Itahashi:2012ut}
points of view for the study of the $\eta'$ mass at finite density.  In
Refs.~\cite{Lee:1996zy,Cohen:1996ng,Jido:2011pq}, 
it has been pointed out theoretically that
the anomaly effect can contribute to the $\eta'$ mass only with
the presence of the spontaneous and/or explicit breaking of chiral
symmetry.  This is because the chiral
singlet gluon current cannot couple to the chiral pseudoscalar mesonic
state without the chiral symmetry breaking.\footnote{For the flavor
SU(2){, in which the strange quark is 
infinitely heavy,} nontrivial topological sectors can
contribute differently to two-point functions of $\eta$ and $\eta'$ even
in the chiral symmetric limit.  Thus $\eta$ and $\eta'$ do not
degenerate in the chiral restoration limit~\cite{Lee:1996zy}.}
\begin{figure}[hbt]
\center
\includegraphics[width=0.85\linewidth]{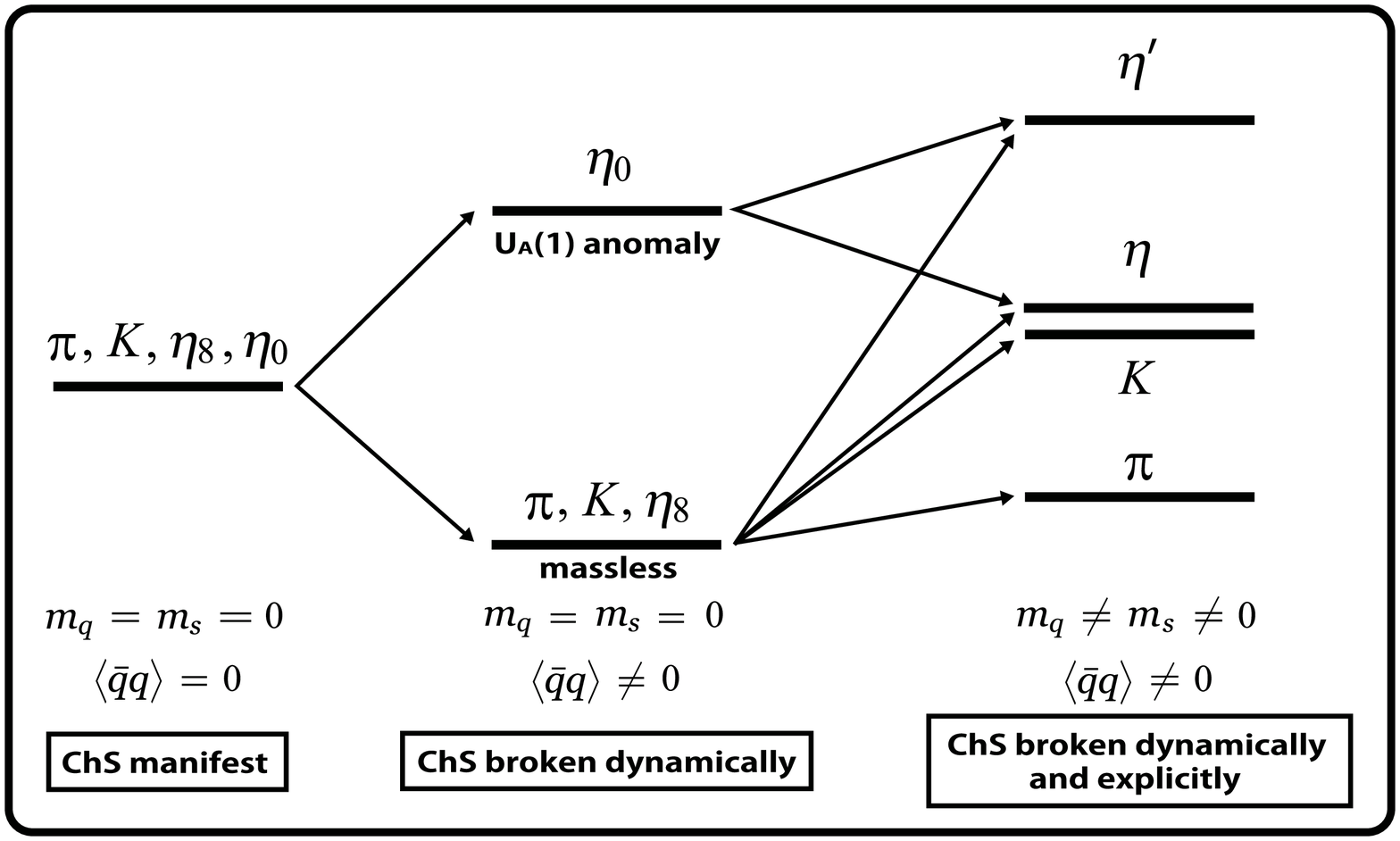}
\caption{
\label{fig:meson_mass}
Light pseudoscalar meson spectrum in the
various patterns of the SU(3) chiral symmetry
breaking. In the left the chiral symmetry is manifest
without explicit nor dynamical breaking. All
the pseudoscalar mesons have a common mass. In the
middle, chiral symmetry is dynamically broken in
the chiral limit. The octet pseudoscalar mesons are
identified as the Nambu-Goldstone bosons associated
with the symmetry breaking. In the right chiral
symmetry is broken dynamically by the quark condensate
and explicitly by finite quark masses.
} 
\end{figure}
Thus, even if density dependence of the U$_{A}$(1) anomaly effect is
irrelevant or negligible, a relatively large mass reduction ($\sim 100$~MeV)
of the $\eta'$ meson 
is expected at nuclear density due to the partial restoration of chiral
symmetry. 
{In Fig.~\ref{fig:meson_mass} we show a schematic view of the
pseudoscalar meson spectra in various chiral symmetry breaking patterns.}

{Meanwhile,} in Ref.~\cite{Itahashi:2012ab,*Itahashi:2012ut}, it has been reported
that the experimental observation of the $\eta'$-nuclear bound states
predicted in Refs.~\cite{Nagahiro:2004qz,Nagahiro:2006dr} is
considered to be possible.
This observation will help us much to understand
the 
$\eta'$ mass generation {mechanism} quantitatively.  In the meson-nucleus bound
system like deeply bound pionic atoms~\cite{Yamazaki:2012PR}, because it is 
guaranteed that the meson inhabits the nucleus, it is unnecessary to
remove in-vacuum contributions from the spectrum.  The fact that the
bound states have definite quantum numbers is favorable to extract
fundamental quantities, since detailed spectroscopy enables us to
investigate selectively the contents of the in-medium meson
self-energy~\cite{Itahashi:1999qb,Yamazaki:2012PR,Suzuki:2002ae}.
Since so far there have been no 
observations of the $\eta'$ meson bound in nuclei, it is
extremely desirable to search for experimental signals of
$\eta'$ bound states in nuclei as a first step to the detailed
investigation of the in-medium $\eta'$ meson properties. 

Thus, in this article, we show the comprehensive {calculation} of the formation
spectra of the $\eta'$ meson-nucleus systems in the $(p,d)$
reaction~\cite{Itahashi:2012ab,*Itahashi:2012ut} based on the latest theoretical
considerations of $\eta'$ property in nucleus~\cite{Jido:2011pq,Nagahiro:2011fi}.
The numerical results shown here are 
important both to give theoretical supports {and predictions} to the planed
experiment in Ref.~\cite{Itahashi:2012ab,*Itahashi:2012ut}{,} and to make it possible to
deduce clearly the $\eta'$
property in nucleus from the experimental data.  

We should stress that
this $\eta'$ mass reduction mechanism has a unique
feature~\cite{Jido:2011pq}.  In 
usual meson-nuclear systems, attractive interactions induced by
many-body effects unavoidably accompany comparably large
absorptions. This is because attractive interaction and absorption
process are originated by the same hadronic many-body effects. This
implies that the bound states have a comparable absorption width with
the level spacing.  In the present case, however, since the suppression
of the U$_{A}$(1) anomaly effect in nuclear medium induces the
attractive interaction to the in-medium $\eta'$ meson, the
influence acts selectively on the $\eta'$ meson and, thus, it
{hardly}
induce{s} inelastic transitions of the $\eta'$ meson into
lighter mesons, although other many-body effects can introduce nuclear
absorptions of the $\eta'$ meson. Consequently the
$\eta'$ meson bound state {is expected to} have a larger binding energy with
a smaller width~\cite{Jido:2011pq}.   This feature is supported by the
theoretical optical potential evaluated in Ref.~\cite{Nagahiro:2011fi}
based on the theoretical $\eta'N$ scattering amplitude~\cite{Oset:2010ub}.

As for other experimental information obtained so far, it has been
reported that a strong reduction of the $\eta'$ 
mass, at least 200 MeV, is necessary to explain the two-pion correlation
in Au + Au collisions at RHIC~\cite{Csorgo:2009pa}.  {In contrast,}
the low-energy $\eta'$ production experiment with $pp$ collision has
suggested a relatively smaller scattering length of the $s$-wave
$\eta'$-proton interaction, $|\mathrm{Re}\ a_{\eta'  p}| <0.8\
\mathrm{fm}$~\cite{Moskal:2000gj} and  $| a_{\eta' p}| \sim 0.1\
\mathrm{fm}$~\cite{Moskal:2000pu}, which corresponds to from several to
tens MeV mass reduction at nuclear saturation density estimated under
the linear density approximation.  
{Transparency ratios for $\eta'$ mesons have been measured for different
nuclei by the CBELSA/TAPS collaboration
\cite{Nanova:2012vw,Nanova:2011jk}.  An absorption width of the $\eta'$
meson at saturation density as small as 15--25 MeV has been found.
Within the experimental uncertainties this width seems to be almost
independent of the $\eta'$ momentum.}
Theoretically,  the
formation spectra of the 
$\eta'$ mesic nuclei are calculated first in
Ref.~\cite{Nagahiro:2004qz}. Nambu--Jona-Lasinio model calculations
suggested around 150 MeV mass reduction of the
$\eta'$ meson at the 
saturation density~\cite{Costa:2002gk,Nagahiro:2006dr}.  
The theoretical model of
Refs. \cite{Oset:2010ub,Nagahiro:2011fi} shows, however,
that the existence of the strong attraction is not consistent with the
latest data of the small scattering length \cite{Moskal:2000pu}.  

{
This article is organized as follows.  In Sec.~\ref{sec:formulation}, we
give the formulation to calculate the formation spectra of the
$\eta'$-nucleus bound states.  The calculated spectra are shown in
Sec.~\ref{sec:result}.  We also show the formation spectra with various
optical potential cases in Appendix.  Finally, we will devote
Sec.~\ref{sec:conclusion} to conclude this article.
}

\section{Formulation of $\eta'$-nucleus bound state formation}
\label{sec:formulation}
{{In this article, we
consider the $(p,d)$ reaction for the formation of the $\eta'$-mesic
nuclei, which can be performed at existing facilities like
GSI~\cite{Itahashi:2012ab,*Itahashi:2012ut}.}
{A missing mass spectroscopy is considered here and it has}
been proved to be a powerful tool for the formation of the meson bound
states~\cite{Toki:1990fc,*Hirenzaki:1991us}.  
In this spectroscopy, one observes only an emitted particle in a final
state, and obtains the double differential cross section 
$d^2\sigma/d\Omega dE$
as a function of the energy of the $\eta'$-nucleus system which is
uniquely determined by the emitted particle energy by means of the
energy conservation law.
}  

To evaluate the formation cross section, we use the Green's function
method~\cite{Morimatsu:1985pf,*[{{\it ibid., }}]Morimatsu:1988dc,Nagahiro:2008rj}. 
In this method, the reaction cross section is assumed to be
separated into the nuclear response function $R(E)$ and the elementary
cross section of the $pn\rightarrow d\eta'$ process with the impulse
approximation:
\begin{equation}
\left(\frac{d^2\sigma}{d\Omega dE}\right)
_{A(p,d)(A-1)\otimes\eta'}
=\left(\frac{d\sigma}{d\Omega}\right)
^{lab}_{n(p,d)\eta'}\times R(E),
\label{eq:impulse}
\end{equation}
where the nuclear response function $R(E)$ is given in terms of the in-medium 
Green's function $G(E)$ as 
\begin{equation}
  R(E) = -\frac{1}{\pi} {\rm Im} \sum_{f} 
\int d{\bf r}d{\bf r'}
{\cal T}_f^{\dagger}({\bf r}) 
G(E; {\bf r}, {\bf r^{\prime}}) {\cal T}_f({\bf r'}) \ \ .
\end{equation}
Here, the summation is inclusively taken over all
possible final states. The 
amplitude ${\cal T}_f$ describes the transition of the incident proton to
a neutron hole and the outgoing deuteron: 
\begin{equation}
  {\cal T}_f({\bf r}) = \chi^{*}_{d}({\bf r}) 
  \left[Y_{l_{\eta'}}^{*}(\hat r)\otimes \psi_{j_{n}}({\bf r})\right]_{JM} \chi_{p}({\bf r})
\label{eq:ampT}
\end{equation} 
with the neutron hole wavefunction $\psi_{j_n}$, 
the distorted waves of proton and the ejected deuteron $\chi_{p}$ and $\chi_{d}$, and
the $\eta'$ angular wavefunction $Y_{l_{\eta'}}(\hat r)$.
For the neutron hole, we use the harmonic oscillator wavefunction for
simplicity. 
The Green's function $G(E)$ contains 
the $\eta'$-nucleus optical potential in the Hamiltonian as 
\begin{equation}
   G(E; {\bf r}, {\bf r^{\prime}}) = \langle n^{-1} | \phi_{\eta'}({\bf r})
\frac{1}{E-H_{\eta'}+i\epsilon} \phi_{\eta'}^{\dagger}({\bf r^{\prime}}) |
n^{-1} \rangle
\label{eq:Green_function}
\end{equation}
where $\phi^{\dagger}_{\eta'}$ is the $\eta'$ creation operator and
$|n^{-1}\rangle$ is the neutron hole state. 
%%%
The elementary cross section in the laboratory frame in
Eq.~(\ref{eq:impulse}) was evaluated to be 30~$\mu$b/sr at the proton
kinetic energy $T_p=2.5$ GeV in
Ref.~\cite{Itahashi:2012ab,*Itahashi:2012ut,Nakayama}. 
%%%
The Green's function $G(E,{\bf r},{\bf r'})$ can be obtained by
solving the Klein-Gordon equation with the appropriate boundary condition. 
Thus, the Green's function represents both the $\eta'$ meson scattering 
states and bound states together with the decay modes
which are expressed in the imaginary part of the potential. 
The imaginary part of the Green's function, or the spectral function, 
represents the coupling strength of the $\eta'$ meson 
to each intermediate state as a function of the energy of the $\eta'$ meson. 
If there are a quasi-bound state of the $\eta'$ meson, the spectral
function has a peak structure at the corresponding energy. 
This can be seen in the formation spectra as a signal of the bound state. 

{
In this article, to discuss the observation feasibilities, we go through
various cases with different optical potentials for the $\eta'$-nucleus
system.  If the mass reduction like expected by the NJL
calculation takes place in nuclear matter, we can translate its effect
into a potential form.  The optical potential $U_{\eta'}(r)$ can be
written as
\begin{equation}
 U_{\eta'}(r)=V(r)+iW(r)
\label{eq:V}
\end{equation}
where $V$ and $W$ denote the real and imaginary parts of the optical
potential, respectively. }
The mass term in the
Klein-Gordon equation for the $\eta'$ meson at finite density can be
written as
\begin{eqnarray}
 m_{\eta'}^2 \rightarrow m_{\eta'}^2(\rho) & = & (m_{\eta'}+\Delta
  m(\rho))^2 \nonumber \\
&\sim & m_{\eta'}^2+2m_{\eta'}\Delta m_{\eta'}(\rho),
\end{eqnarray}
where $m_{\eta'}$ is the mass of the $\eta'$ meson in vacuum and
$m_{\eta'}(\rho)$ the mass at finite density $\rho$.  
The mass shift
$\Delta m_{\eta'}(\rho)$ is defined as 
$\Delta m_{\eta'}(\rho)=m_{\eta'}(\rho)-m_{\eta'}$.  Thus, we can
interpret the 
mass shift $\Delta m_{\eta'}(\rho)$ as the strength of the real part of
the optical potential 
\begin{equation}
 V(r)=\Delta m_{\eta'}(\rho_0)\frac{\rho(r)}{\rho_0}\ 
{\equiv V_0\frac{\rho(r)}{\rho_0}},
\end{equation}
in the Klein-Gordon equation
using the mass shift at normal saturation density $\rho_0$. 
%
%For the $\eta'$-nucleus optical potential, we simply assume an
%empirical form as,
%\begin{equation}
% V_{\eta'}=(V_0+iW_0)\frac{\rho(r)}{\rho_0},
%\label{eq:V}
%\end{equation}
%with the nuclear density distribution $\rho(r)$ and the normal
%saturation density $\rho_0=0.17$ fm$^{-3}$.  The mass term in the
%Klein-Gordon equation for the $\eta'$ meson at finite density can be
%written as
%\begin{eqnarray}
% m_{\eta'}^2 \rightarrow m_{\eta'}^2(\rho) & = & (m_{\eta'}+\Delta
%  m(\rho))^2 \nonumber \\
%&\sim & m_{\eta'}^2+2m_{\eta'}\Delta m_{\eta'}(\rho),
%\end{eqnarray}
%where $m_{\eta'}$ is the mass of the $\eta'$ meson in vacuum and
%$m_{\eta'}(\rho)$ the mass at finite density $\rho$.  The mass shift
%$\Delta m_{\eta'}(\rho)$ is defined as 
%$\Delta m_{\eta'}(\rho)=m_{\eta'}(\rho)-m_{\eta'}$.  Thus, we can
%interpret the 
%mass shift $\Delta m_{\eta'}(\rho)$ as the strength of the real part of
%the optical potential in the Klein-Gordon equation
%\begin{equation}
% V_0=\Delta m_{\eta'}(\rho_0)\ ,
%\end{equation}
%using the mass shift at normal saturation density $\rho_0$.  
%
%{The theoretical optical potentials obtained in
%Ref.~\cite{Nagahiro:2011fi}, which are obtained by a theoretical
%$\eta'N$ scattering length~\cite{Oset:2010ub} using the standard many
%body theory, are also considered to calculate the $\eta'$-nucleus states
%and the formation cross sections. } 
%
Here we
assume the nuclear density distribution $\rho(r)$ to be of an empirical
Woods-Saxon form as
\begin{equation}
 \rho(r)=\frac{\rho_N}{1+\exp(\frac{r-R}{a})},
\end{equation}
where $R=1.18 A^{\frac{1}{3}}-0.48$ fm, $a=0.5$ fm with the nuclear
mass number $A$, and $\rho_N$ a normalization factor such that
$\int d^3 r \rho(r)=A$.
{In the following sections, we show the
$(p,d)$ spectra with {the potential depth from} $V_0=$ 0 {to}
$-200$ MeV and $W_0=-5$ {to} $-20$ MeV to 
discuss the observation feasibility, where $W_0$ is the strength of the
imaginary part of the optical potential at $\rho_0$.
}

{
Alternatively, {we also use the} theoretical optical potentials for
the $\eta'$-nucleus 
system {obtained} in Ref.~\cite{Nagahiro:2011fi} by {imposing
several} theoretical 
$\eta'N$ scattering length{s}~\cite{Oset:2010ub} {and} using the
standard many 
body theory.  There the two-body absorption of the $\eta'$ meson in a
nucleus together with the one-body absorption has been evaluated so that
we can decompose the spectra into the different final states by using
the Green's function method as discussed below. 

We obtain the in-medium Green's function by solving
the Klein-Gordon equation with the optical potential $U_{\eta'}$ in
Eq.~(\ref{eq:V}) with the appropriate boundary condition and use it to
evaluate the nuclear response function 
$R(E)$ in Eq.~(\ref{eq:impulse}).

We estimate the flux loss of the injected proton and the ejected
deuteron due to the elastic and quasi-elastic scattering and/or
absorption processes by {the} target {and} daughter {nuclei.} 
{To estimate the attenuation probabilities,}
we approximate the distorted waves of the incoming proton $\chi_p$ 
and the outgoing deuteron $\chi_d$ as,
\begin{equation}
\chi_d^*({\bm r})\chi_p({\bm r}) = \exp\left[i{\bm q}\cdot{\bm r}\right]
F({\bm r}),
\label{eq:F}
\end{equation}
with the momentum transfer between proton and deuteron ${\bm q}={\bm
p}_p-{\bm p}_d$ and
{the distortion factor $F({\bm r})$ evaluated by}
\begin{multline}
F({\bm r})=\exp\left[-\frac{1}{2}\sigma_{p N}\int_{-\infty}^z
dz' \rho_A(z',b)\right.
\\
\left.
-\frac{1}{2}\sigma_{dN}\int_z^\infty dz'
\rho_{A-1}(z',b)
\right]. \label{eq:distfactor}
\end{multline}
%%%
{Here $\sigma_{p N}$ and $\sigma_{dN}$ are the proton-nucleon and 
deuteron-nucleon total cross sections, respectively, which contain 
both  the elastic and {inelastic} processes. 
The values of the total cross sections 
are taken from Ref.~\cite{PDG08}. 
}
$\rho_A(z,b)$ is the density distribution function for the nucleus with
the mass number $A$ in  cylindrical coordinates.

The calculation of the formation spectra is done separately
{for each} subcomponent of the $\eta'$-mesic nuclei labeled by
$(n\ell_{j})^{-1}_{n}\otimes 
\ell_{\eta'}$, which means a configuration of a neutron-hole
in the $\ell$ orbit  with the total spin $j$ and the principal quantum 
number $n$ in the daughter nucleus
and an $\eta'$ meson in the $\ell_{\eta'}$ orbit.
The total formation spectra are obtained by summing up these
subcomponents taking into account the difference of the separation
energies for the different neutron-hole states.

{The energy of the emitted deuteron determines the energy of the
$\eta'$-nucleus system uniquely.}
We show the 
calculated spectra 
{as functions of the excitation energy $E_{\rm
ex}-E_0$ defined as
\begin{equation}
 E_{\rm ex} - E_0 = - B_{\eta'} + [S_n(j_n)-S_n({\rm ground})],
\end{equation}
where $B_{\eta'}$ is the $\eta'$ binding energy and $S_n(j_n)$ the
neutron separation energy from the neutron single-particle level
$j_n$.  $S_n({\rm ground})$ indicates the separation energy from the
neutron level corresponding to the ground state of the daughter
nucleus.  $E_0$ is the $\eta'$ production threshold 
energy.

The widths of the hole states are taken into account in the present
calculation.  The width of the neutron-hole states in 
{$^{11}$C have been estimated to be $\Gamma((0s_{1/2})^{-1})=12.1$
MeV for the excited state and
$\Gamma((0p_{3/2})^{-1})=0$ MeV for the ground state
by using the data in Ref.~\cite{11Chole}.
As for $^{39}$Ca,  we use $\Gamma=7.7$ MeV $((1s_{1/2})^{-1})$, 3.7 MeV
$((0d_{5/2})^{-1})$, 21.6 MeV $((0p_{3/2,1/2})^{-1})$, and 30.6 MeV
$((0s_{1/2})^{-1})$ estimated from the data in
Ref.~\cite{Nakamura:1974zz}, 
considering the width of the ground state $(0d_{3/2})^{-1}$ to be 0 and
assuming the same widths for neutron-hole states as those of proton
holes.  }

%$^{39}$Ca have
%been estimated to be $\Gamma=7.7$ MeV $((1s_{1/2})^{-1})$, 3.7 MeV
%$((0d_{5/2})^{-1})$, 21.6 MeV $((0p_{3/2,1/2})^{-1})$, and 30.6 MeV
%$((0s_{1/2})^{-1})$ from the data in Ref.~\cite{Nakamura:1974zz},
%considering the width of the ground state $(0d_{3/2})^{-1}$ to be 0 and
%assuming the same widths for neutron-hole states as those of proton
%holes.  
%As for $^{11}$C, we have used the data in Ref.~\cite{11Chole}
%and the widths are $\Gamma((0s_{1/2})^{-1})=12.1$ MeV for the
%excited state and
%$\Gamma((0p_{3/2})^{-1})=0$ MeV for the ground state.
%%

In the Green's function
method~\cite{Morimatsu:1985pf,*Morimatsu:1988dc}, one can 
separately calculate each contribution to the spectrum coming from
the different $\eta'$ processes. 
On the prescription of Ref.~\cite{Morimatsu:1985pf,*Morimatsu:1988dc},
we rewrite equivalently the imaginary part of the Green's function of
$\eta'$ as  
\begin{equation}
{\rm Im}G = (1+G^\dagger U_{\eta'}^\dagger){\rm Im}G_0 (1+U_{\eta'} G) + 
G^\dagger{\rm Im}U_{\eta'} G\ , 
\label{eq:Green}
\end{equation}
where $G$ and $G_0$ denote the full and free Green's functions for $\eta'$
and $U_{\eta'}$ is the $\eta'$-nucleus optical potential.
{We abbreviate the integral symbols in Eq.~(\ref{eq:Green}).}
The first term of the right-hand-side of
Eq.~(\ref{eq:Green}) represents the contribution from the escape 
$\eta'$ from the daughter nucleus and the second term describes the 
conversion process caused by the $\eta'$ absorption into the nucleus. 
By evaluating only the conversion part, we obtain spectra associated 
with decays (or absorptions) of the $\eta'$ mesons in 
the nucleus, which correspond to the coincident measurements in real
experiments.
\begin{figure}[hbt]
\center
\includegraphics[width=0.85\linewidth]{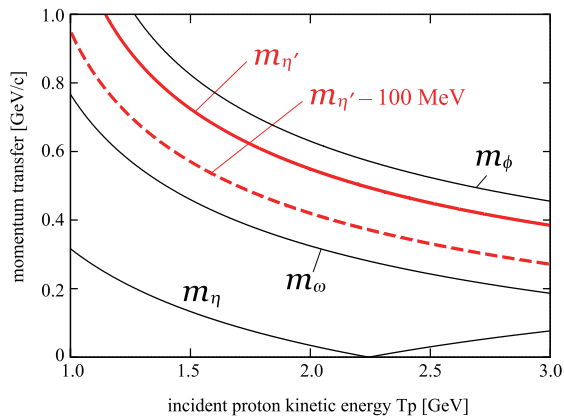}
\caption{(color online) Momentum transfer of the $^{12}$C$(p,d)$ reactions as functions of the
 incident proton kinetic 
 energy $T_p$. The thick solid and dashed lines correspond to the $\eta'$ meson
 production with the binding energy of 0 MeV and 100 MeV.  
Thin solid lines correspond to the $\eta$, $\omega$, and $\phi$ meson
 productions with the binding energy of 0 MeV, as indicated in the
 figure. 
\label{fig:mom_trans}
} 
\end{figure}

\section{Numerical Results}
\label{sec:result}

First,
we show in Fig.~\ref{fig:mom_trans} the momentum transfer of the $(p,d)$
reactions for the formation of the $\eta'$ meson.  We also show those for
the $\eta$, $\omega$, and $\phi$ meson production cases.  Those mesons,
which have relatively closer masses to that of $\eta'$, can contribute
to the $(p,d)$ spectrum in the same energy
region~\cite{Nagahiro:2006dr}. We find that the recoilless condition can
be satisfied only for the $\eta$ production case in this energy region.
For the $\eta'$ 
production case, the recoilless condition is never satisfied even for
the $\eta'$ bound states with the binding energy of 100 MeV. 
The momentum transfer at $T_p=2.5$ GeV, which is the energy considered
in Ref.~\cite{Itahashi:2012ab,*Itahashi:2012ut}, is around 400 -- 500 MeV/$c$ and thus,
various contributions of $(n\ell_j)_n^{-1}\otimes\ell_{\eta'}$ will
contribute to the $(p,d)$ spectrum.

We show the calculated {formation spectra of $\eta'$-nucleus system
for the $^{12}$C target case with}
the potential strength
$(V_0,W_0)=-(0,10)$ and $-(100,10)$ MeV cases in Fig.~\ref{fig:V0}.  As
we can see from the figure, the existence of the attractive interaction
and bound states can be seen as the peak structures in the $(p,d)$
spectrum.  {We find that there is a clear difference between the
cases with attractive and non-attractive potentials.}

\begin{figure}[hbt]
\center
\includegraphics[width=1\linewidth]{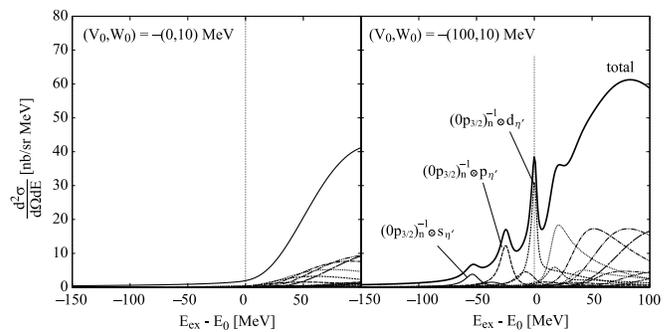}
\caption{Calculated spectrum of $^{12}$C$(p,d)$$^{11}$C$\otimes\eta'$
 reaction for the formation of the $\eta'$-nucleus systems
with the proton kinetic energy $T_p=2.5$ GeV and the deuteron
 angle $\theta_d=0^\circ$ as a function of the excited energy 
$E_{\rm ex}$. $E_0$ is the $\eta'$ production threshold.  The depth of
 the $\eta'$-nucleus optical potential is (a) $(V_0,W_0)=-(0,10)$ MeV, and
(b) $(V_0,W_0)=-(100,10)$ MeV.
The thick solid line shows the total spectrum and dashed lines indicate
 subcomponents.  The neutron-hole states are indicated as
 $(n\ell_j)_n^{-1}$ and the $\eta'$ states as $\ell_{\eta'}$.  
\label{fig:V0}
}
\end{figure}

\begin{figure*}[hbt]
\center
\includegraphics[width=0.95\linewidth]{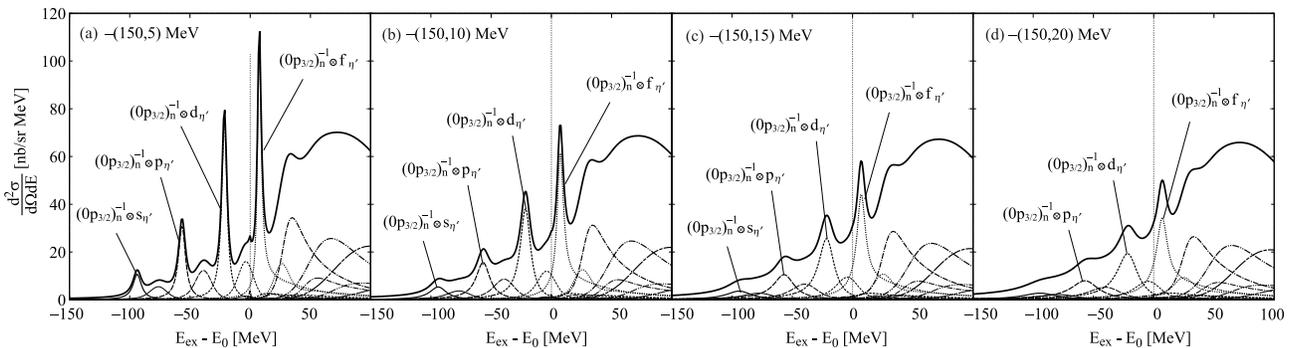}
\caption{Calculated spectra of $^{12}$C$(p,d)$$^{11}$C$\otimes\eta'$
 reaction 
 for the formation of the $\eta'$-nucleus systems
with the proton kinetic energy $T_p=2.5$ GeV and the deuteron
 angle $\theta_d=0^\circ$ as functions of the excited energy 
$E_{\rm ex}$. $E_0$ is the $\eta'$ production threshold.  The 
 $\eta'$-nucleus optical potential are
(a) $(V_0,W_0)=-(150,5)$ MeV, (b) $-(150,10)$ MeV, 
(c) $-(150,15)$ MeV, and (d) $-(150,20)$ MeV.
The thick solid lines show the total spectra and dashed lines indicate
 subcomponents.  The neutron-hole states are indicated as
 $(n\ell_j)_n^{-1}$ and the $\eta'$ states as $\ell_{\eta'}$.  
\label{fig:V150}
}
\end{figure*}

\begin{figure}[hbt]
\center
\includegraphics[width=0.95\linewidth]{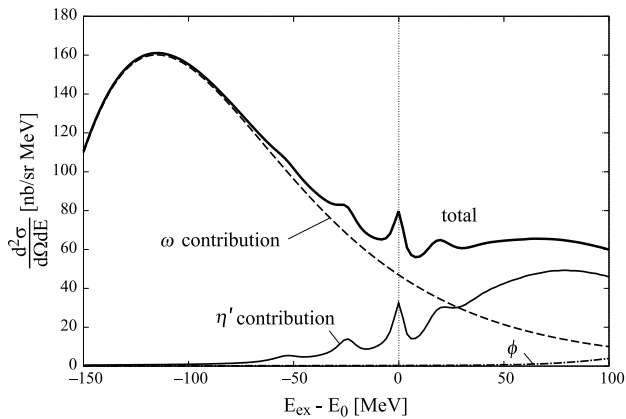}
\caption{Calculated spectra of $^{12}$C$(p,d)$$^{11}$C$\otimes\eta'$,
$^{12}$C$(p,d)$$^{11}$C$\otimes\omega$, and $^{12}$C$(p,d)$$^{11}$C$\otimes\phi$
 reactions 
 for the formation of the meson-nucleus systems
with the proton kinetic energy {$T_p=2.6$ GeV} and the deuteron
 angle $\theta_d=0^\circ$ as functions of the excited energy 
$E_{\rm ex}$. $E_0$ is the $\eta'$ production threshold. 
 The 
 $\eta'$-nucleus optical potential $(V_0,W_0)=-(100,10)$ MeV,
 $\omega$-nucleus optical potential $(V_0,W_0)=-(-42.8,19.5)$ MeV~\cite{Lutz:2001mi,*Lutz:2001mierr},
and the $\phi$-nucleus optical potential is $(V_0,W_0)=-(30,10)$
 MeV~\cite{YamagataSekihara:2010rb,Muto:2005za,Cabrera:2002hc,Hatsuda:1991ez}.  
{The thin solid line shows the $\eta'$ production, and the dashed
 and dot-dashed lines indicate the $\omega$ and $\phi$ meson {productions.}
 {The} contributions of mesonic states with partial waves up to
 $\ell=6$ for each meson {are included} in the calculation.
 }
\label{fig:green_phi}
}
\end{figure}

{In Fig.~\ref{fig:V150}, we show the effects of the absorption
interaction by varying the strength of the imaginary part $W_0$ of the
optical potential.  We can see the clear peaks corresponding to bound
states in the $s$, $p$ and $d$ states, although the width of each peak
becomes wider as $W_0$ is increased.  We also find that there are peak
structures in the $f_{\eta'}$-wave component just above the threshold
($E_{\rm ex}-E_0=0$) owing to the so-called threshold enhancement.
While 
there are no bound state in $f_{\eta'}$ state of $\eta'$, the
attractive $\eta'$-nucleus interaction pulls this low energy scattering
wave of $\eta'$ closer to the daughter nucleus enhancing its overlap
with the nucleon wavefunctions and consequently producing a larger cross
section.  Therefore, we can consider this enhancement to give an
indication of the attractive $\eta'$-nucleus interaction if
observed.   We find that, even in {a} large imaginary case of
$-(150,20)$~MeV, we can see a clear peak corresponding to this threshold
enhancement indicating the attractive nature of the $\eta'$-nucleus
optical potential.  In Appendix, we show various cases of the strength
of the real optical potentials to see the {experimental feasibility}
systematically.  We find that, in the weak
attraction $V_0=-50$~MeV case, we cannot see peak structures with larger
absorption $|W_0| { \gtrsim } 15$~MeV, while in the strong attraction
{$|V_0|\gtrsim 100$~MeV} case we can see clear peaks even in the large
absorption $W_0=-20$~MeV, which corresponds to the absorption width
$\Gamma=40$~MeV.
}

{The contributions from other meson production processes are shown in
Fig.~\ref{fig:green_phi}.  }
{Because we are considering the inclusive reaction in this article, the
{productions of}
other mesons having close masses also contribute {to} the
spectra of the
$(p,d)$ reaction {in addition to} the
formation of the $\eta'$-mesic nuclei.  Here, we take the contributions
from the $\omega$ and $\phi$ mesons into account in
Fig.~\ref{fig:green_phi}.
The incident proton kinetic energy is set to be $T_p=2.6$~GeV.
The $\phi$-nucleus interaction is taken as
$V_\phi=-(30+10i)\rho(r)/\rho_0$~MeV that corresponding the 3\% mass
reduction of the $\phi$ meson at normal saturation density.  The
imaginary part of the optical potential has been estimated by using the
chiral unitary approach and $W_0=-10$~MeV is used
here {as in Ref.}~\cite{YamagataSekihara:2010rb}.  The 
elementary cross section of $pn\rightarrow d\phi$ is estimated as
$\left(\dfrac{d\sigma}{d\Omega}\right)^{\rm lab.}=$13.5 $\mu$b/sr by using the
experimental data~\cite{Maeda:2006wv}.  As we can see the figure, the
contribution from the $\phi$ meson is negligibly small owing to the
large momentum transfer for the $\phi$ meson production.  }

{
In contrast to the $\phi$ meson, we find that the $\omega$ meson
production gives larger contribution to the $\eta'$ bound region.
Although it is still unknown whether the $\omega$-nucleus interaction
is attractive or repulsive, we consider the case that the
$\omega$-nucleus optical potential is repulsive as
$V_\omega=-(-42.8+19.5i)\rho(r)/\rho_0$~MeV~\cite{Lutz:2001mi,Nagahiro:2005gf}
because in the repulsive case the quasi-free $\omega$ contribution above
the $\omega$ production threshold is enhanced and then it overlaps the
$\eta'$ bound region~\cite{Nagahiro:2006dr}.  The elementary cross
section of $pn\rightarrow d\omega$ in laboratory frame with
$T_p=2.6$~GeV is estimated as 27~$\mu$b/sr by using the experimental
data~\cite{Barsov:2003zc}.  As shown in Fig.~\ref{fig:green_phi},
although the contribution of the quasi-free $\omega$ is large, the
strength of the tail around the $\eta'$ threshold, where we can see the
clear peak of $\eta'$, is of the same order of the $\eta'$ signal and
does not have any structure there.  Therefore, we can expect to observe
the peak structure of $\eta'$ even if there is a {quasi-elastic}
$\omega$ contribution. 
}

\begin{figure*}[hbt]
\center
\includegraphics[width=0.95\linewidth]{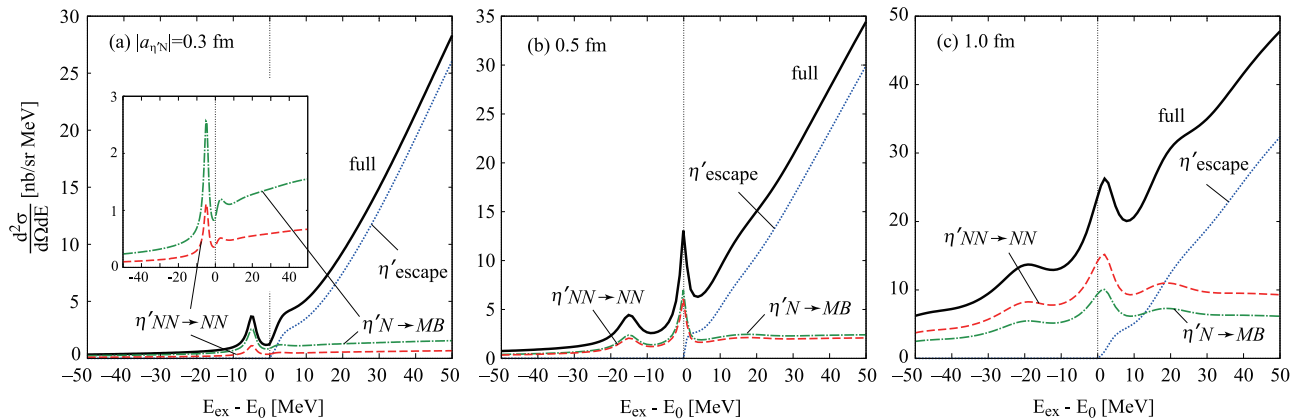}
\caption{(color online) Calculated spectra of $^{12}$C$(p,d)$$^{11}$C$\otimes\eta'$ 
 reaction for the formation of the $\eta'$-nucleus systems
with the proton kinetic energy $T_p=2.5$ GeV and the deuteron
 angle $\theta_d=0^\circ$ as functions of the excited energy 
$E_{\rm ex}$. $E_0$ is the $\eta'$ production threshold. 
 The 
 $\eta'$-nucleus optical potential are evaluated in
 Ref.~\cite{Nagahiro:2011fi}, which correspond with the $\eta'$
 scattering length $|a_{\eta'p}|=$ (a) 0.3, (b) 0.5, and (c) 1.0 fm, respectively.
The thick solid lines show the total spectra and dashed lines show
 subcomponents as indicated in the figure.  The inset figure in
 panel-(a) shows the structure of the subcomponents in closeup.
\label{fig:oset_pot}
}
\end{figure*}

In Fig.~\ref{fig:oset_pot}, we show the results calculated with the
theoretical optical potentials in Ref.~\cite{Nagahiro:2011fi}.  The
potential parameters used here are corresponding to the $\eta'N$
scattering length of $|a_{\eta'N}|=$0.3, 0.5, and 1.0 fm,
respectively~\cite{Nagahiro:2011fi,Oset:2010ub}.  
{In these figures, we show only the $\eta'$ contribution as in
Figs.~\ref{fig:V0} and \ref{fig:V150}.}
{As shown in
Fig.~\ref{fig:oset_pot}, we decompose the total spectrum for each case
into three parts; the contribution from the $\eta'$-escape process and
two conversion parts of the one-body absorption $\eta'N\rightarrow MB$
(where $M$ denotes a meson and $B$ a baryon which {are} included in the
coupled-channel calculation in Ref.~\cite{Oset:2010ub}) and two-body
absorption $\eta'NN\rightarrow NN$.  The spectra with coincident
observations of a nucleon-pair associated with the two-body absorption of
$\eta'$ are shown by the dashed-lines and those of a meson-baryon
(mostly $\eta N$ or $\pi N$) pair by the dot-dashed lines.
{These} conversion spectra shown in Fig.~\ref{fig:oset_pot} give useful
information for coincident measurements of the decay particles from
the $\eta'$-bound states, which could reduce the large
background estimated in Ref.~\cite{Itahashi:2012ab,*Itahashi:2012ut}. 
}

We show the results with the heavier $^{40}$Ca target in
Fig.~\ref{fig:40Ca}.  {In some cases, a larger target is more
suitable because there are more bound states.  In the case of the
$\eta'$-mesic nuclei formation by the $(p,d)$ reaction, however, the
bound state peaks overlap each other because of the smaller level
spacing than the $^{12}$C case.  Therefore, we conclude that a smaller
target like $^{12}$C is better suited for the $\eta'$-mesic nuclei
formation. 
}

\begin{figure}[hbt]
\center
\includegraphics[width=0.95\linewidth]{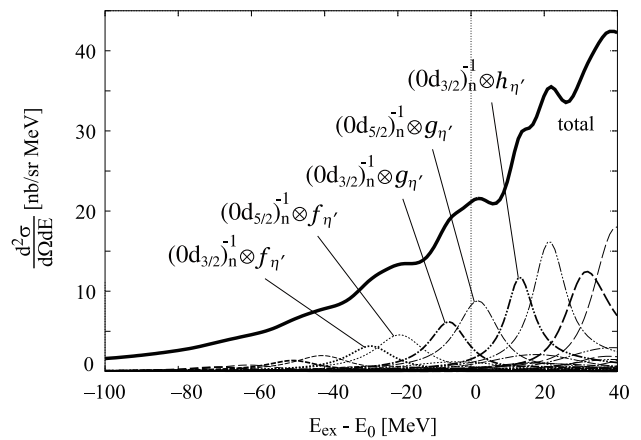}
\caption{Calculated spectra of $^{40}$Ca$(p,d)$$^{39}$Ca$\otimes\eta'$
 reaction for the formation of the $\eta'$-nucleus systems
with the proton kinetic energy $T_p=2.5$ GeV and the deuteron
 angle $\theta_d=0^\circ$ as functions of the excited energy 
$E_{\rm ex}$. $E_0$ is the $\eta'$ production threshold.  The 
 $\eta'$-nucleus optical potential is $(V_0,W_0)=-(100,10)$ MeV.
The thick solid lines show the total spectra and dashed lines indicate
 subcomponents.  The neutron-hole states are indicated as
 $(n\ell_j)_n^{-1}$ and the $\eta'$ states as $\ell_{\eta'}$.  
\label{fig:40Ca}
}
\end{figure}

{
\section{Conclusions}
\label{sec:conclusion}
We have calculated the formation spectra of the $\eta'(958)$-nucleus
systems {for} the $(p,d)$ reaction.  The kinetic energy of the incident
proton beam is set to be $T_p=2.5$--$2.6$~GeV, {which can} be reached at
existing facilities like GSI~\cite{Itahashi:2012ab,*Itahashi:2012ut}.
We have shown the numerical results for various strengths of the
$\eta'$-nucleus optical potentials as {from $V_0=0$ to $-200$ MeV
and $W_0=-5$ to $-20$ MeV}
as well as no attraction case as $(V_0,W_0)=-(0,10)$~MeV.  We find that,
in the strong attraction case $|V_0|\gtrsim 100$~MeV, that is {the}
expected strength {of the attraction}
by the NJL calculation, we can see clear peaks around the $\eta'$
production threshold even with the large absorption {case as} $W_0=-20$~MeV. 
In some cases, the peaks around the threshold {do} not
{indicate the existence of the}
bound
state{s} but the so-called threshold enhancement{s} which
{are} also consequence{s} of the attractive nature of the $\eta'$-nucleus
interaction.  The robustness of the appearance of the peak structure around
the threshold for an attractive interaction, which is independent on the
detail of the {model parameters within the range of the present
consideration}, is an interesting and important finding of this 
study.  We conclude that we can see clear signal of the possible
attractive potential of the $\eta'$-nucleus system by the $(p,d)$
reaction with $^{12}$C target.  The conversion spectra accompanied by
different absorption process of $\eta'$ in nucleus are discussed, which
give useful information to the coincident measurements of the decay
particles from the $\eta'$ bound states.
}

{We also have looked at the contributions from other meson
productions whose mass is close to that of $\eta'$.  Although they must
be {one of the} sources of the background, the contributions are
{almost} structure-less and
then don't {disturb} the peak observation of $\eta'$.  We have discuss the heavier
target nucleus case as well, because it is often said that a larger
target is more suitable to make a bound state.  We find that, however,
in the present case a relatively light nucleus as $^{12}$C likely to
have not many but a few bound states is suited to observe peaks in the
formation spectra.}

{
So far, the relatively small scattering length of the $\eta'$-nucleon has
been reported~\cite{Moskal:2000gj,Moskal:2000pu} although its sign is
still unknown.  If such a small scattering length is a consequence of
a weak attraction, it is difficult to observe the $\eta'$-nucleus bound
states by using the proposed method here.  In such a case, we have to
develop an advanced understanding of such a small $\eta'$-nucleon
interaction from the microscopic and fundamental point of view.  
}

{{In contrast}, the transparency ratios of the $\eta'$ meson
have been measured by the CBELSA/TAPS
collaboration~\cite{Nanova:2012vw,Nanova:2011jk} and suggest remarkably
small absorption width of the $\eta'$ meson in nuclear medium as compared
to other meson like $\eta$ or $\omega$, which is consistent
with a scenario for the fate of $\eta'$ in finite density discussed in
Ref.~\cite{Jido:2011pq}. 
}

{
In any case, an experimental searching for bound $\eta'$ in nuclei would
provide an important information on the properties of $\eta'$.  
We believe that the present theoretical results are much important for
such experimental activities to obtain the deeper insight of the meson mass
spectrum.
}

\section*{Acknowledgement}
{
We appreciate the useful discussions with E.~Oset, A.~Ramos, V.~Metag,
and M.~Nanova. This work is supported by the Grant-in-Aid for Scientific
Research (Nos.~24105707 (H.~N.), {22740161 (D.~J.)}, and 24540274 (S.~H.))
in Japan. {A part of this work was done under Yukawa International Project for 
Quark-Hadron Sciences (YIPQS).}}

\section*{Appendix}

In this appendix, we show the calculated {
$^{12}$C$(p,d)^{11}$C$\otimes\eta'$} spectra {at $T_p=2.5$~GeV} with
various 
combinations of the potential strength with the range of
$V_0$ {from} $-50$ {to} $-200$ MeV and $W_0=-5$ {to} $-20$
MeV
{in Fig.~\ref{fig:V50-200}.}
\begin{figure*}[hbt]
\center
\includegraphics[width=0.95\linewidth]{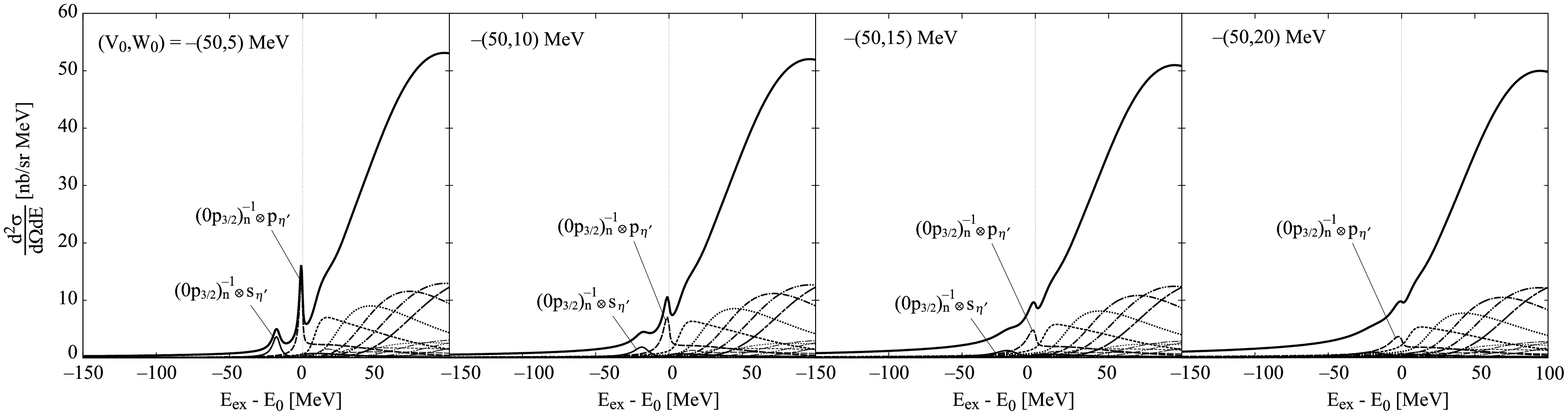}
\includegraphics[width=0.95\linewidth]{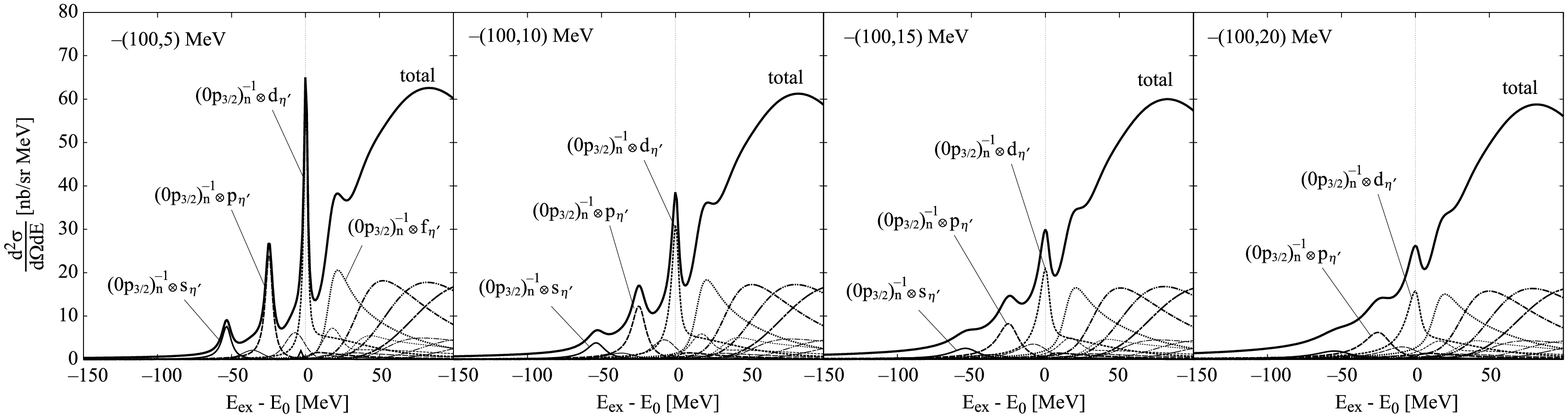}
\includegraphics[width=0.95\linewidth]{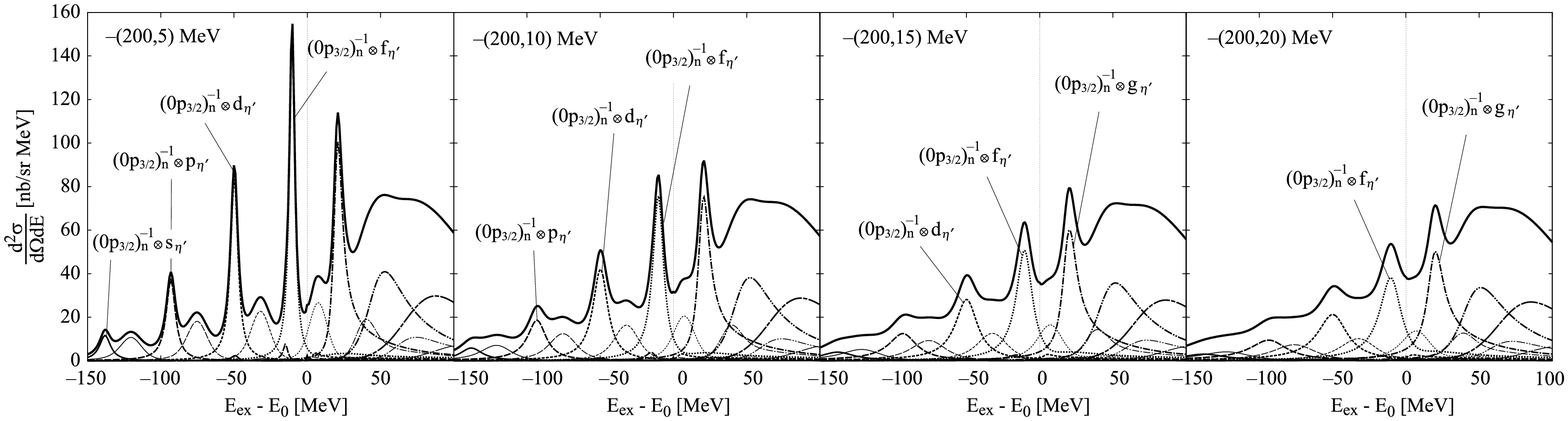}
\caption{Calculated spectra of $^{12}$C$(p,d)$$^{11}$C$\otimes\eta'$
 reaction for the formation of the $\eta'$-nucleus systems
with the proton kinetic energy $T_p=2.5$ GeV and the deuteron
 angle $\theta_d=0^\circ$ as functions of the excited energy 
$E_{\rm ex}$. $E_0$ is the $\eta'$ production threshold. 
Various combinations of the potential strength are considered within the
 range of $V_0=-50$ -- $-200$ MeV and $W_0=-5$ -- $-20$ MeV as indicated
 in the figure.
The thick solid lines show the total spectra and dashed lines indicate
 subcomponents.  The neutron-hole states are indicated as
 $(n\ell_j)_n^{-1}$ and the $\eta'$ states as $\ell_{\eta'}$.  
\label{fig:V50-200}
}
\end{figure*}

\bibliography{etaprime}
\end{document}